\documentstyle[preprint,aps]{revtex}

\begin{document}

\title{Gravitational Lensing Statistics as a Probe of Dark Energy}

\author{Zong-Hong Zhu\thanks{e-mail address: zhuzh@class1.bao.ac.cn}
	}

\address{Beijing Astronomical Observatory,
                Chinese Academy of Sciences, Beijing 100012, China\\
        National Astronomical Observatories,
                Chinese Academy of Sciences, Beijing 100012, China
        }

\maketitle

\begin{abstract}

By using the comoving distance, we derive an analytic expression for the 
optical depth of gravitational lensing, which depends on the redshift to
the source and the cosmological model characterized by the cosmic mass
density parameter $\Omega_m$, the dark energy density parameter $\Omega_x$ 
and its equation of state $\omega_x = p_x/\rho_x$.
It is shown that, the larger the dark energy density is and the more negative 
its pressure is, the higher the gravitational lensing probability is. 
This fact can provide an independent constraint for dark energy.\\

\end{abstract}

\section{Introduction}

The standard cosmological model is based on three cornerstones: the Hubble
expansion, the Cosmic Microwave Background Radiation(CMBR) and the primordial
Big Bang Nucleosynthesis.
Now these three kinds of observations, such as Hubble's relation for
fifty-some Ia type supernovae(SNeIa) out to redshifts of nearly 
one\cite{per98rie98}, the 
anisotropy of the CMBR\cite{lin98} and the deuterium abundance measured 
in four high redshift hydrogen clouds seen in absorption against distant
quasars\cite{bur98}(combined with baryon fraction in galaxy clusters
from X-ray data\cite{whi93}), have made a strong case for the existence
of a nearly uniform component of dark energy with negative presure.
It seems that determining the amount and nature of the dark energy
is emerging as one of the most important challenges in cosmology.

One apparent plausible candidate for the dark energy is the cosmological 
constant or vacuum energy density\cite{wei89,kra95}. 
The possibility of a nonzero 
cosmological constant $\Lambda$ has been advocated and then discarded several 
times in the past for theoretical and observational reasons\cite{coh98sah99}.
Due to this checked history and the difficulty in understanding the
observed $\Lambda$ in the framework of modern  quantum field 
theory\cite{wei89}, Most physicists and astronomers believe that the 
cosmological constant should be zero because of some unknown physics.
Other candidates for the dark energy include: a frustrated network of
topological defects (such as strings or walls)\cite{defect} and an
evolving scalar field (referred to by some as quintessence)\cite{quint}.
As shown in literatures, it is difficult to discriminate well against 
these different possibilities either by the SNIa data alone\cite{gar98} or
only by the CMBR data\cite{hue99}. This led some authors to consider
the combination of the SNIa measurements with the anisotropy of 
CMBR\cite{tur97} or the large scale structure\cite{per99}.

In this paper, an independent means for probing the amount 
and nature of dark energy is proposed. 
It relies on the gravitational lensing statistics, which has been 
shown to be an efficient tool for determining the cosmological 
parameters\cite{lensing}. Some authors even have given the general
expressions for the optical depth and mean image seperation in general
Friedman-Robertson-Walker (FRW) cosmological models\cite{analytic}.
But these expressions are complicated and thereby hard to apply in practice.  
By using the comoving distance we derive an analytic and simple expression 
for the optical depth of gravitational lensing that depends on the redshift to
the source and the cosmological model characterized by the cosmic mass
density parameter $\Omega_m$, the dark energy density parameter $\Omega_x$ 
and its equation of state $\omega_x = p_x/\rho_x$.
It is shown that for a flat universe the lensing probability is very
sensitive to $\Omega_x$ and $\omega_x$, and hence provides an independent
probe for the dark energy.

\section{Kinematics with dark energy component}

We assume a homogeneous and isotropic universe with Robertson-Walker
metric (in the $c = 1$ unit):
\begin{equation}
ds^2 = -dt^2 + R^2(t) \left[ d\chi^2 +
f^2(\chi)(d \theta^2 + \sin^2\theta d\phi^2)\right],
\label{metric}
\end{equation}
where $f(\chi) = \chi$ for a flat universe ($k=0$), $f(\chi) = \sin\chi$
for a closed universe ($k= + 1$), and $f(\chi) = \sinh\chi$ for an open
universe ($k= - 1$).
Defining the scale factor $a(t) = R(t)/R_0$,  $a=1$ today, and
the Friedmann equation takes the form
\begin{equation}
\left(\frac{\dot{a}}{a} \right)^2 =
\frac{8 \pi G}{3}\sum_i \rho_i - \frac{k}{a^2 R_0^2}
\label{friedmann}
\end{equation}
where $i$ includes all components of matter and energy in the universe.
If the effective equation of state for the $i$-component is parametrized
as $\omega_i = p_i/\rho_i$, its density scales as $\rho_i \propto a^{-n_i}$
where $n_i = 3(1+\omega_i)$. 
For instance, nonrelativistic matter scales
as $\rho_m \propto a^{-3}$ while relativistic matter, such as radiation,
changes as $\rho_r \propto a^{-4}$, and vacuum energy(cosmological constant)
is invariant ($\rho_{\Lambda} \equiv \Lambda/(8 \pi G) \propto a^0$)
as the universe expands.
For the purpose of studying gravitational lensing statistics, we need 
only consider two important components: one is the
nonrelativistic matter ($\rho_m$), the other is a nearly uniform dark
energy ($\rho_x, \omega_x = p_x/\rho_x$) with a constant $\omega_x$.
In order to avoid interfering with structure formation, the dark energy
component must be less important in the past than matter although it may
dominates the universe today. So $\omega_x$ and hence the pressure of
dark energy must be negative\cite{tur98}.
Defining
\begin{equation}
\Omega_m =  \frac{8 \pi G }{3 H_0^2} \rho_{m0}, \,\,\,\,
\Omega_x =  \frac{8 \pi G }{3 H_0^2} \rho_{x0}, \,\,\,\,
\Omega_k =  \frac{-k}{R_0^2 H_0^2}.
\label{omega}
\end{equation}
where $H_0$ is the Hubble constant, $\rho_{m0}$ and $\rho_{x0}$ are 
the nonrelativistic matter density and the dark energy density at
present respectively,
then Eq.\ref{friedmann} becomes
\begin{equation}
\frac{1}{a }\frac{da}{dt}  = H_0 \left( \frac{\Omega_x}{a^{3(1+\omega_x)}}
		+ \frac{\Omega_m}{a^3} +\frac{\Omega_k}{a^2} 
			\right)^{1/2}.
\label{dadt}
\end{equation}
Note that we have a relation, $1=\Omega_x + \Omega_m + \Omega_k$.
The distance a light ray travels can be calculated as following.
Light rays follow null geodesics where 
$ds^2 =0$ so that $dt^2 = R_0^2 a^2 d\chi^2$.
It has been shown that the natural cosmological distance for the analysis
of gravitational lensing statistics is the comoving distance\cite{zhu98}.
With Eq.\ref{dadt} and the relation $a(t) = (1+z)^{-1}$, the comoving
distance is
\begin{equation}
\chi = \left\{ \begin{array}{ll}
		\int_0^z \frac{dz}{\sqrt{\Omega_x (1+z)^{3(1+\omega_x)}
				+ (1-\Omega_x) (1+z)^3}}    & (k=0)\\
			& \\
		\left|\Omega_k\right|^{1/2} \int_0^{z}
				\frac{dz}{E(z)}           & (k=\pm 1)\\
		\end{array}
	\right. 
\label{comoving}
\end{equation}
$$
E(z) \equiv \sqrt{\Omega_x (1+z)^{3(1+\omega_x)} + \Omega_m (1+z)^3
			+\Omega_k (1+z)^2 }     .
$$
The comoving volume $dV$ of the shell $d\chi$ at $\chi$ reads 
\begin{equation}
dV = 4\pi R^3(t_0) f^2(\chi)d\chi \, .
\label{volume}
\end{equation}

\section{Optical depth of gravitational lensing in general FRW cosmologies}

Now, let's calculate the optical depth of gravitational lensing by all
galaxies with different luminosities and redshifts in the universe.
Following Turner et al.\cite{tog84},  we model the mass density profile of 
a galaxy matter as the singular isothermal sphere (SIS) parameterized by
its velocity dispersion $\sigma$. 
The dimensionless cross-section of multiple images for a point source 
located at $z_s$ produced by a single SIS galaxy at $z_d$ 
is\cite{analytic,zhu98}
\begin{equation}
\hat{\sigma} = 16{\pi}^3 \sigma^4
		\left[\frac{f(\chi_s - \chi_d)}{f(\chi_s)}\right]^2 \,.
\label{sigma}
\end{equation}
By accumulating the contributions of various galaxies,
the total dimensionless cross-section by galaxies at redshifts ranging 
from 0 to $z_s$ for distant sources like quasars at $z_s$ is
\begin{equation}
\hat{\Sigma}(z_s) = 4\pi \times F \times T(z_s), \,\,\,\,\,
			F = \sum_{i = E, S0, S} F_i \,.
\label{sumsigma}
\end{equation}
The parameter $F_i$, which depends only on the intrinsic and statistical 
properties of galaxies in the universe,  represents the effectiveness of 
the $i$-th morphological type of galaxies in producing multiple 
images\cite{tog84}. Due to its uncertainties discussed widely in
literatures\cite{lensing,analytic,zhu98}, we treat $F$ as a normalized
factor hereon.
The $z_s$ dependent factor $T(z_s)$  is
\begin{equation}
T(z_s) = (H_0 R_0)^3 \int_0^{\chi_s} {\left[ \frac{f(\chi_s - \chi_d)}
			{f(\chi_s)} \right]}^2 f^2(\chi_d) d\chi_d
\label{Tzs}
\end{equation}
After some algebra calculation, we get an analytically simple expression
for the optical depth (probability) of gravitational lensing for a point
source at $z_s$ in general FRW cosmologies with dark energy
\begin{equation}
p = p(z_s; \Omega_m, \Omega_x, \omega_x) \\
= \left\{ \begin{array}{ll}
		\frac{F}{30} \chi^3_s &  (k=0)\\
			& \\
		\frac{F}{\left|\Omega_k\right|^\frac{3}{2}} 
			\left[\frac{1}{8}(1 + 3 \cot^2\chi_s) \chi_s -
			\frac{3}{8}\cot\chi_s\right] & (k=+1)\\
			& \\
		\frac{F}{\left|\Omega_k\right|^\frac{3}{2}}
			\left[\frac{1}{8}(-1 + 3 \coth^2\chi_s) \chi_s -
			\frac{3}{8}\coth\chi_s\right] & (k=-1)\\
		\end{array}
	\right. \\
\label{probability}
\end{equation}
where $\Omega_k = 1 - \Omega_m - \Omega_x$ and $\chi_s$ can be calculated 
through Eq.~\ref{comoving}.
In the above derivations, the comoving number density of galaxies is 
assumed to be constant. However, this may not hold true for the
realistic situation. If the galaxy evolution depends on cosmological models,
one can hardly divide the optical depth into $F$-term and $T$-term\cite{mao94}.
Fortunately, by using the galaxy merging model proposed by Broadhurst 
et al.\cite{merging} which can successfully account for both 
the redshift distribution and the number counts of galaxies at
optical and near-infrared wavelengths, Zhu and Wu\cite{zhu97} has shown 
that the galaxy merging doesn't affect the optical depth of lensing.

\section{Results and discussions}

Knowing the analytic expressions for the lensing optical depth in general
FRW cosmologies, it is easy to demonstrate how the lensing probability
depends sensitively on the amount and nature of dark energy.
For sake of simplicity,  we now concentrate on flat universe models,
which is strongly supported by various observational 
evidences\cite{per98rie98,lin98,bur98} and prefered by 
the inflationary scenario.
Fig.1 shows how the probability $p$ depends on a vacuum energy component
(cosmological constant) for a flat universe.
In our calculations, the source has been set at $z_s= 1, 2, 3, 4, 5$
respectively.
Indeed, the probability $p$ depends sensitively on the cosmological constant,
this is why astronomers take it as a promising means to determine $\Lambda$
\cite{lensing}.

Generally, the dark energy is parameterized by 
($\Omega_x, \omega_x = p_x/\rho_x$).
Therefore the optical depth of gravitational lensing depends on the amount 
of dark energy $\Omega_x$ as well as its equation of state $\omega_x$.
The larger the dark energy amount is, the higher the gravitational lensing 
probability is; the more negative the dark energy pressure is, the higher
the optical depth is (see Fig.2). 
This fact can provide an independent constraint for dark energy, as a 
complement to other methods.
Fig.2 is the contour plots for gravitational lensing probability (normalized 
to the parameter $F$) in the ($\omega_x, \Omega_x$) plane.
Along each contour, the dark energy model is degenerated for a lensing
optical depth. Fortunately, this degeneracy can be resolved when
combining the lensing data with SNIa and CMBR measurements\cite{wag99}.
Thus the gravitational lensing statistics can serve as an efficient
but independent tool for probing the dark energy.

In summary, by using the comoving distance we derive an analytic and simple 
expression for the optical depth of gravitational lensing that depends on the 
redshift to the source and the cosmological model characterized by the cosmic 
mass density $\Omega_m$, the dark energy density $\Omega_x$ and its equation
of state $\omega_x = p_x/\rho_x$.
It is shown that, for a flat universe, the lensing probability is very
sensitive to $\Omega_x$ and $\omega_x$ and hence provides an independent
probe for the dark energy.\\

This work was supported by the National Natural Science Foundation of China,
under Grant No. 19903002.

\begin{figure*}[htbp]
\caption{The lensing probability (in unit of the effectiveness of galaxies
        in producing multiple images $F$) as a function of the cosmological
        constant for different source redshifts.  
        The universe is assumed to be flat.
        }
\end{figure*}

\begin{figure*}[htbp]
\caption{The contour plots for lensing probability (in unit of the
effectiveness of galaxies in producing multiple images $F$) in the
($\omega_x, \Omega_x$) plane, where the universe is assumed to be flat
and the source is set at $z_s = 3$.}
\end{figure*}

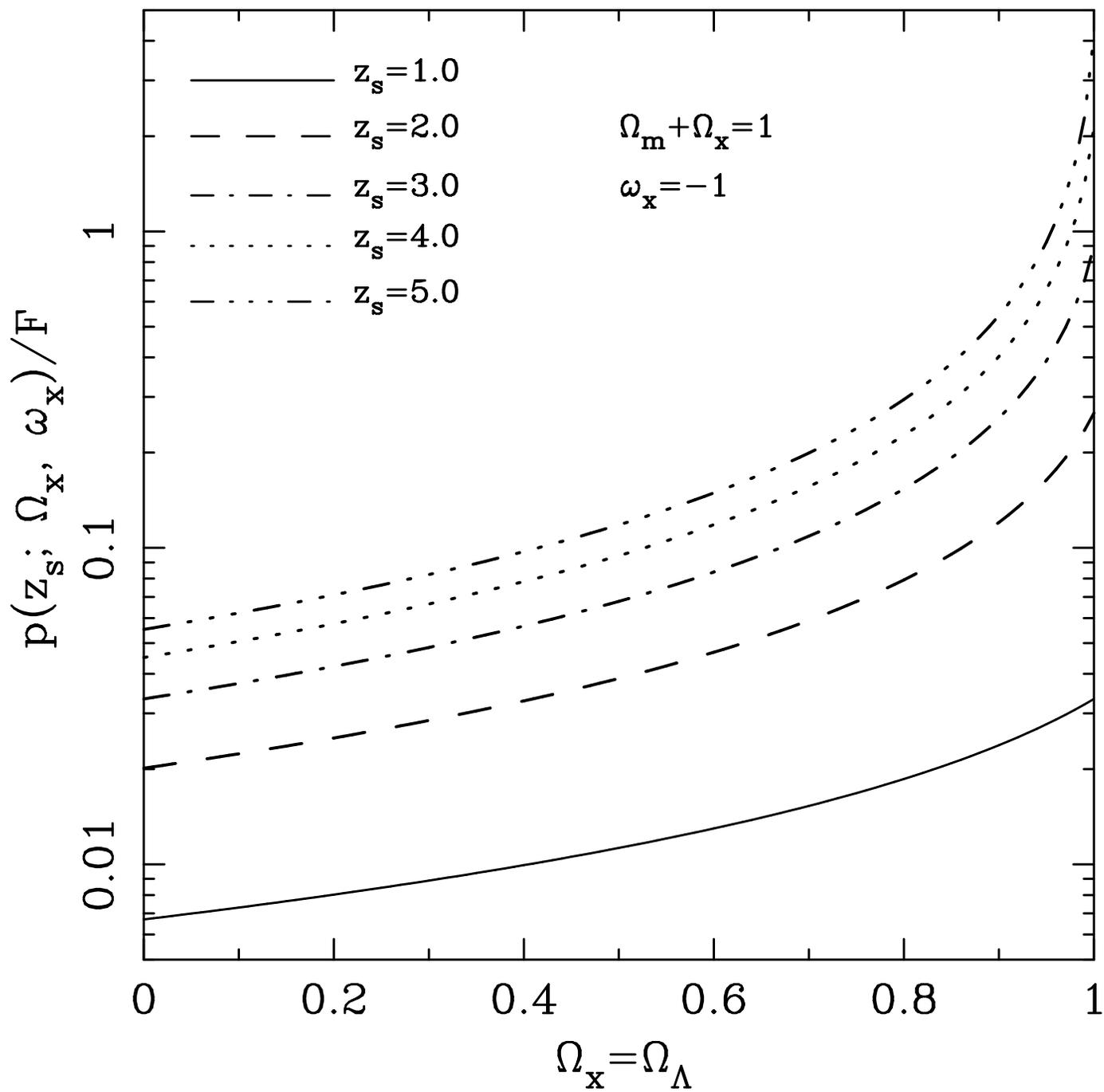
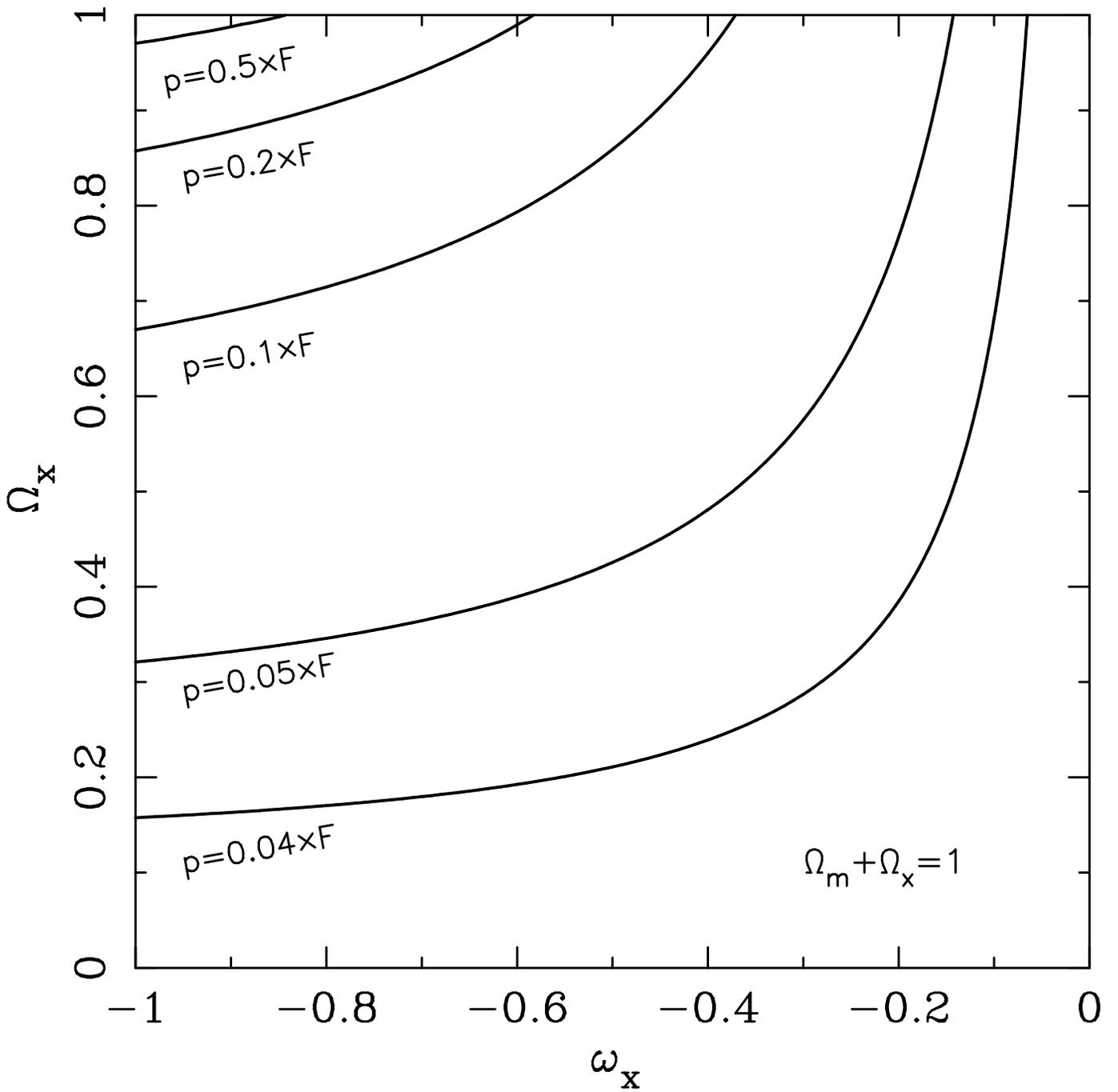
\end{document}